\documentstyle[12pt]{article}

\begin{document}
\baselineskip 5mm

\makeatletter
\def\eqnarray{%
 \stepcounter{equation}%
 \let\@currentlabel=\theequation
 \global\@eqnswtrue
 \global\@eqcnt\z@
 \tabskip\@centering
 \let\\=\@eqncr
 $$\halign to \displaywidth\bgroup\@eqnsel\hskip\@centering
 $\displaystyle\tabskip\z@{##}$&\global\@eqcnt\@ne
 \hfil$\displaystyle{{}##{}}$\hfil
 &\global\@eqcnt\tw@$\displaystyle\tabskip\z@{##}$\hfil
 \tabskip\@centering&\llap{##}\tabskip\z@\cr}
\makeatother

\title{The Berkovits Method for Conformally Invariant Non-linear $\sigma$-Models on $G/H$   }

\author{Shogo Aoyama\thanks{e-mail: spsaoya@ipc.shizuoka.ac.jp} \\
       Department of Physics \\
              Shizuoka University \\
                Ohya 836, Shizuoka  \\
                 Japan}
                 
\maketitle
 
\vspace{2cm}

\begin{abstract}

We discuss 2-dimmensional non-linear $\sigma$-models on the K\"ahler manifold $G/H$ in the first order formalisim. Using the Berkovits method we explicitly construct  the $G$-symmetry currents and primaries, when $G/H$ are irreducible.  It  is a  variant of the Wakimoto realization of the affine Lie algebra using   a  particular reducible K\"ahler manifold $G/U(1)^r$ with $r$ the rank of $G$.

\end{abstract}

\vspace{5cm}



\newpage

The Berkovits formalism for the superstring\cite{Berk} is expected to be a new way which overcomes the long-standing problem of the Rammond-Neveu-Schwarz(RNS) and Green-Schwarz(GS) formalisms. The right-moving contribution of the GS superstring action was put in the linealized form
\begin{eqnarray}
S=\int d^2z({1\over 2}\partial x^m\bar\partial x_m + \rho_\alpha\bar\partial \theta^\alpha).   \label{berkaction}
\end{eqnarray}
After the Wick rotation the $SO(10)$ currents are given by $M^{mn}={1\over 2}\rho\gamma^{mn}\theta$ in this  formalism, while they are given by $\hat M^{mn}=\psi^m\psi^n$ in the RNS formalism. There is a crucial difference between the OPE's of both currents. To cope with this discrepancy Berkovits added a $bc$ system\cite{Marti} taking  the form 
\begin{eqnarray}
S'= \int d^2z (v^{ab}\bar\partial u_{ab} +\beta\bar\partial \gamma).  \label{modaction} 
\end{eqnarray}
Here $u_{ab}$ are the coordinates parametrizing the coset space $SO(10)/U(5)$ and belong to ${\bf 10}$ of $U(5)$ and  $\gamma$ is a bosonic ghost.  $v^{ab}$ and $\beta$ are their canonical conjugate momenta for the respective quantity. He constructed new $SO(10)$ currents $N^{mn}$ by using this $bc$ system, so that the modified $SO(10)$ currents $M'^{mn}={1\over 2}\rho\gamma^{mn}\theta +N^{mn}$  have the same OPE's as  $\hat M^{mn}=\psi^m\psi^n$. Moreover the combined theory given by $S+S'$ is free of conformal anomaly. Quantization of the superstring is done  by studying the BRST cohomology. To this end  the pure spinor $\lambda^\alpha \in {\bf 16}$ of $SO(10)$  
satisfying 
\begin{eqnarray}
\lambda\gamma^m\lambda=0  \label{pure}
\end{eqnarray}
plays an essential role.  Decomposing ${\bf 16}$ under $U(5)$ as ${\bf 1} + {\bf 10} + {\bf 5}$ we can solve this equation\cite{Berk, kaza} by using  the fields of the $bc$ system (\ref{modaction}) as
\begin{eqnarray}
\lambda^\alpha =\left(
\begin{array}{c}
  \gamma  \\
\gamma u_{ab} \\
   -{1\over 8}\gamma\epsilon^{abcde}u_{bc}u_{de} 
\end{array}
\right).   \label{pureprim}
\end{eqnarray}
The  OPE's $N^{mn}(z)\lambda^\alpha(w)$ yield the correct $SO(10)$ algebra
 only if the bosonic ghosts are fermionized as $\beta =\partial \xi e^{-\varphi}$ and $\gamma = \eta e^{\varphi}$. 

In short, the point of the Berkovits formalism is to construct the  currents $N^{mn}$ of weight $1$ and the primaries $\lambda^\alpha$ of weight $0$, belonging to ${\bf 45}$ and ${\bf 16}$ of $SO(10)$ respectively, by using the $bc$ system which manifests the $U(5)$ symmetry 
alone. 
We note that $SO(10)/U(5)$ is a K\"ahler coset space. The construction may be generalized with  the general K\"ahler  coset space $G/H$, though our concern deviates from the central issue of the formalism about the superstring.
It is a variant of  the free field realization of the WZWN model with $G$ symmetry, which is called the Wakimoto realization of the affine Lie algebra of $G$.  The Wakimoto realization itself was well studied by many people\cite{wakimoto,Peter}. Nonetheless in this letter we pursue the study using the Berkovits method  for the following reasons. Firstly in \cite{wakimoto,Peter} the affine Lie algebra of $G$ is realized by using the $bc$ system 
\begin{eqnarray}
S'' = \int d^2z (\sum_{\alpha\in \Delta_+} p_\alpha\bar\partial q^\alpha + \sum_{i=1}^{r} \varphi^i ),     \label{S''} 
\end{eqnarray}
in which $\Delta_+$ denotes the set of positive roots for the Lie algebra and $r$ is the rank of $G$. 
 The fields $q^\alpha$ parametrize the K\"ahler coset space $G/U(1)^r$. 
 Instead we use the $bc$ system (\ref{modaction}) 
in an extended form so as to parametrize the general K\"ahler coset space $G/H$
 Secondly in \cite{wakimoto,Peter} the explicit formulae for the $G$-symmetry currents were  given  for  those corresponding to the positve or negative simple roots, but for other currents they were yet implicit.  Thirdly  the G-symmetry primaries  were not discussed  in \cite{wakimoto}. In \cite{Peter} they were discussed, but the construction was based on  $G/U(1)^r$ and  not so explicit as (\ref{pureprim}). 
The final reason of the study is that we consider the $bc$ system (\ref{modaction}) or (\ref{S''})  as a non-linear $\sigma$-model on $G/H$  formulated in the first order. 
Then  we can see a close relationship between  the $G$-symmetry currents and the Killing potential which exists for the general K\"ahler coset space $G/H$.  The K\"ahler geomery  underlying in the the affine Lie algebra of $G$ is different  depending on which subgroup $H$ is taken.

The aim of this letter is to give an explicit and simple construction  of the $G$-symmetry currents and primaries, by parametrizing  the K\"ahler coset space $G/H$  in the case where  it is irreducible. 
As for the $G$-symmetry primaries  we are interested in those of weight $0$. 
It suffices to find them in the  fundamental representation (or the spinorial representation for $G=SO(n)$). Then those  in any other representation can be constructed  by tensoring them.  We would like to stress on the fact that the resulting primaries satisfy the $G$-symmetry algebra only if the bosonic ghosts are ferminized according to Berkovits. 
At the end of the letter it will be pointed out that 
the K\"ahler geometry  of the affine Lie algebra  is useful to study  the non-commutative geometry.

We start with a brief summary of  the K\"ahler geometry. 
Consider a real $2N$-dimentional symplectic manifold $\cal M$ with local coordinates $(x^1,x^2\cdots,x^{2N})$. The line-element and the symplectic  $2$-form  respectively  given by 
\begin{eqnarray}
d s^2 = {1\over 2} g_{ij}dx^i dx^j,   \quad\quad\quad
\omega = {1\over 2}\omega_{ij}dx^i \wedge dx^j.    \label{metric}
\end{eqnarray}
 We write the world-sheet action of a non-linear $\sigma$-model on $\cal M$ as
\begin{eqnarray}
S={1\over 2}\int d^2\xi [\eta^{ab}g_{ij} + \varepsilon^{ab}\omega_{ij}]\partial_a x^i\partial_b x^j.    \label{action}
\end{eqnarray}
$\cal M$ is a K\"ahler manifold if the $2$-form $\omega$ is closed and there exists  a complex structure $J_i^j$ such that $J_i^j J_j^k =-\delta_i^k$ and $\omega_{ij}=g_{ik}J_j^k$. We locally set $J_i^j$ to be
$$
J_i^j = \left(
\begin{array}{cc}
  -i\delta^\alpha_\beta & 0 \\
    &    \\
 0 & i\delta^{\bar\alpha}_{\bar\beta}
\end{array}
\right).
$$
Then (\ref{metric}) and (\ref{action}) are reduced to 
\begin{eqnarray}
ds^2 = g_{\alpha\bar\beta}dx^\alpha dx^{\bar\beta}, \quad\quad\quad 
\omega = ig_{\alpha\bar \beta}dq^\alpha\wedge dq^{\bar\beta},  \label{metric'}
\end{eqnarray} 
and 
\begin{eqnarray}
S=\int d^2 z g_{\alpha\bar \beta}\partial  q^{\bar\beta}\bar\partial q^\alpha,   \label{action'}
\end{eqnarray}
in which  $q^\alpha $ and $q^{\bar\alpha}$  are complex coordinates obtained by  complexifying $x^i$  by the projectors $(1\pm J)_i^j$ respectively. The closure of $\omega$ given  by (\ref{metric'})  implies the existence of a K\"ahler potential $K$ such that 
$$
g_{\alpha\bar\beta}={\partial^2 K\over \partial q^\alpha\partial 
q^{\bar\beta}}.
$$

If the K\"ahler manifold $\cal M$ is a coset space $G/H$, 
there exists  a set of holomorphic Killing vectors $R^{A\alpha}$ satisfying 
\begin{eqnarray}
{\cal L}_{R^A} R^{B\alpha} &=&  f^{ABC}R^{C\alpha}  \label{Lie},  \\
{\cal L}_{R^A} g_{\alpha\bar \beta} &=& 0.   \label{killcon} 
\end{eqnarray} 
Here ${\cal L}_{R^A}$ is the Lie-derivative with respect to $R^{A\alpha}$ and  $f^{ABC}$ are the structure constants of the symmetry group $G$. 
 Owing to (\ref{killcon}) the action (\ref{action'}) is invariant under $\delta q^\alpha =\epsilon^A R^{A\alpha}$  and $\delta q^{\bar\alpha} =\epsilon^A R^{A\bar\alpha}$ with infinitesimal global parameters $\epsilon^A$ of the $G$-symmetry. But it is not conformally invariant. This model  may be put  in the first order formalism as
\begin{eqnarray}
S=\int d^2 z [p_\alpha\bar\partial q^\alpha + c.c.]  \label{action''},
\end{eqnarray}
by setting  $g_{\alpha\bar \beta}\partial  q^{\bar\beta}$ to be a world-sheet vector $p_\alpha$. Then the action (\ref{action''}) is conformally invariant. It has also the $G$-symmetry under  $\delta q^\alpha =\epsilon^A R^{A\alpha}$
  together with 
$$
\delta p_\alpha = -\varepsilon^A p_\beta R^{A\beta}_{\ \ , \alpha} \Big(\equiv 
-\varepsilon^A p_\beta {\partial R^{A\beta}\over \partial q^\alpha} \Big).
$$
The Noether currents for the $G$-symmetry take the form 
\begin{eqnarray}
J^A =p_\alpha R^{A\alpha}, \quad\quad\quad 
\bar J^A =  p_\alpha R^{A\bar\alpha}.   \label{currents}
\end{eqnarray}
 Owing to (\ref{Lie}) they transform  as the adjoint representation of $G$:
$$
\delta J^A = \varepsilon^A f^{ABC}J^C,
$$
by the $G$-symmetry transformations $\delta q^\alpha$ and $\delta p_\alpha$ above mentioned. This classical argument no longer holds at the quantum level. Namely, if  the $G$-symmetry  is correctly realized at the quantum level,  with the free field OPE
\begin{eqnarray}
 p_\alpha(z)q^\beta(w) \sim \delta_\alpha^\beta {1\over  z-w} \label{pq},
\end{eqnarray}
one should check that 
\begin{eqnarray}
J^A(z)J^B(w) \sim {f^{ABC}J^C(w)\over  z-w} + {kg^{AB}\over (z-w)^2}
 + O\Big((z-w)^{-2} \Big). \label{OPE}
\end{eqnarray}
Here   $k$ is some constant and $g^{AB}$  is the Killing metric of the symmetry group $G$. But the OPE of the currents (\ref{currents}) takes the form 
\begin{eqnarray}
 J^A(z)J^B(w) \sim {f^{ABC}\over z-w} + {R^{A\alpha}_{\ \ ,\beta}(z)R^{B\beta}_{\ \ , \alpha}(w)\over (z-w)^2}
 + O\Big( (z-w)^{-2} \Big), \nonumber
\end{eqnarray}
with the simplified notation $R^{A\alpha}(z)\equiv R^{A\alpha}(q(z))$. 
In general  we have neither 
\begin{eqnarray}
\lim_{z\rightarrow w}R^{A\alpha}_{\ \ ,\beta}(z)R^{B\beta}_{\ \ , \alpha}(w)\ne  k\delta^{AB} \quad{\rm nor}\quad 
\lim_{z\rightarrow w}R^{A\alpha}_{\ \ ,\beta\gamma}(z)\partial q^\gamma (z)R^{B\beta}_{\ \ , \alpha}(w) \ne  0,  \nonumber 
\end{eqnarray}
so that the $G$-symmetry is broken at the quantum level.
To recover the $G$-symmetry we use the Berkovits method. Namely we generalize the action (\ref{action''}) introducing the bosonic ghosts fields $\beta$ and $\gamma$
\begin{eqnarray}
S=\int d^2 z [p_\alpha\bar\partial q^\alpha +  \beta\bar\partial \gamma +c.c.].   \nonumber
\end{eqnarray}
We modify the $G$-symmetry currents by adding terms of weight $(1,0)$ as
\begin{eqnarray}
J^A = p_\alpha R^{A\alpha} + F^A\beta\gamma +G^A_{\ \ ,\alpha}  \partial q^\alpha.  \label{Gcurr}
\end{eqnarray}
Here $F^A$ and $G^A$ are holomorphic functions of $q^\alpha$. The question is whether they can be determined so that the modified currents satisfy the algebra (\ref{OPE}). Lets us check it 
using (\ref{pq}) and  
$$
\beta(z)\gamma(w)\sim {1\over z-w}.
$$ 
We then find that 
\begin{eqnarray}
J^A(z)J^B(w) \sim  {\Lambda^{AB}(w)\over z-w} + {\Theta^{AB}(z,w)\over (z-w)^2} + O\Big( (z-w)^{-2} \Big), \nonumber
\end{eqnarray}
in which
\begin{eqnarray}
\Lambda^{AB}(z) &=&  f^{ABC}p_\alpha R^{C\alpha} +  R^{A\alpha}F^B_{\ ,\alpha}- R^{B\alpha}F^A_{\ ,\alpha}\ , \nonumber    \\
\Theta^{AB}(z,w) &=& R^{A\alpha}_{\ \ ,\beta}(z)R^{B\beta}_{\ \ , \alpha}(w) +F^A(z)F^B(w)   \nonumber\\
&+& \{R^{A\alpha}(z)G^B_{\  ,\alpha}(w)+ G^A_{\  ,\alpha}(z)R^{B\alpha}(w)\}.
\nonumber
\end{eqnarray}
The condition  for this to satisfy (\ref{OPE}) is 
\begin{eqnarray}
R^{A\alpha}F^B_{\ ,\alpha}- R^{B\alpha}F^A_{\ ,\alpha}&=& f^{ABC}F^A, \hspace{3cm} \label{F}\\
\lim_{z\rightarrow w} \Theta^{AB}(z,w) &=& kg^{AB},  \label{G}\\
 \lim_{z\rightarrow w} {\partial \Theta^{AB} (z,w)\over \partial q(z)^\alpha} &=& f^{ABC} G^C_{\ , \alpha}(w).   \label{G'}
\end{eqnarray}
We may identify the holomorphic functions $F^A$ with those appearing in the Lie-variation of the K\"ahler potential 
\begin{eqnarray}
{\cal L}_{R^A} K = F^A + c.c.,    \label{KF}
\end{eqnarray}
modulo a multiplicative constant. Such functions indeed satisfy the condition (\ref{F}) for the K\"ahler coset space in general\cite{bagger}. As for  the other holomorphic functions $G^A$, we propose that 
\begin{eqnarray}
G^A_{\ ,\alpha} \propto \delta_\alpha^A.     \label{G''}
\end{eqnarray}
We now show that the conditions (\ref{G}) and (\ref{G'}) are also satisfied  when the K\"ahler coset space  $G/H$ is irreducible.

The irreducibile K\"ahler coset space $G/H$ is defined as follows. The generators of $G$  are decompsed as
$$
\{X^\alpha, \bar X_\alpha, H^i, Y \},
$$
in which $X^\alpha$ and their  conjugates $\bar X_\alpha$ are coset generators and $Y$ is a $U(1)$ generator. Then $X^\alpha$ ($\bar X_\alpha$) belong to an irreducible representation  under the subgroup  $H$ generated by $H^i$ and $Y$. Such a K\"ahler coset space is called the hermitian symmetric  space and  is characterized by the Lie algebra of the form
\begin{eqnarray}
[ X^\alpha, \bar X_\beta ] &=&  t(\Sigma^i)^\alpha_{\ \beta} H^i +
 s \delta^\alpha_\beta Y,  \quad\quad  [ X^\alpha, X^\beta ] \ =\  0,  \label{geneLie}   \nonumber\\
 \quad [ H^i, X^\alpha ] &=& (\Sigma^i)^\alpha_{\ \beta} X^\beta,\quad\quad [ Y, X^\alpha ]\ =\  X^\alpha, \quad c.c.,   \label{eq 36'}
\end{eqnarray}
with some constants $t$ and $s$ depending on the representation of $G$. 
The local coordinates of the coset space  $q^\alpha$ and $q^{\bar\alpha}$  correspond to the generators $X^\alpha$ and $\bar X_\alpha$ respectively.  From now on we change the notation of  $q^{\bar\alpha}$ as $\bar q_{\alpha}$
 in accordance with that  of $\bar X_\alpha$. Therefore raisng or lowering the tensor indices should be done by writing the metrics $g_\alpha^{\ \beta}$ or $(g^{-1})_\alpha^{\ \beta}$ explicitly. 
Simple algebra gives 
\begin{eqnarray}
[X^\alpha, [X^\beta, \bar X_\gamma ] ] = -\{t (\Sigma^i)^\alpha_{\ \gamma}(\Sigma^i)^\beta_{\ \delta} + s \delta^\alpha_\gamma \delta^\beta_\delta)\} X^\delta 
  \equiv  M^{\alpha\beta}_{\gamma\delta} X^\delta.  \label{eq 37}
\end{eqnarray}
The quantity $M_{\alpha\gamma}^{\beta\delta} $ plays a key role in the method and has a remarkable property. It is summarized by the statement that
$$
M^{\alpha_1\beta_1}_{\gamma_0\gamma_1}M^{\alpha_2\beta_2}_{\gamma_2\beta_1}\cdots M^{\alpha_{n-1}\beta_{n-1}}_{\gamma_{n-1}\beta_{n-2}}M^{\alpha_n\beta_n}_{\gamma_n\beta_{n-1}}
$$
is completely symmetric in the indices $(\alpha_1,\alpha_2,\cdots,\alpha_n,\beta_n)$, whenenver it is completely symmetrized in the indices $(\gamma_0,\gamma_1,\cdots,\gamma_n)$, and vice versa. For the case of $n=1$ we have 
$
M^{\alpha\beta}_{\gamma\delta}=M^{\beta\alpha}_{\gamma\delta} =M^{\alpha\beta}_{\delta\gamma}
$.
With this quantity 
the Killing vectors $R^{A \alpha}$  are given\cite{MA1} by
 \begin{eqnarray}
R_\gamma^{\ \alpha} &=& i\delta_\gamma^\alpha, \quad\quad 
R^{\gamma\alpha} \ =\ {i\over 2}M^{\gamma\alpha}_{\beta\delta} q^\beta q^\delta,  \nonumber  \\
R^{i\alpha} &=&  i(\Sigma^i)^\alpha_{\ \beta} q^\beta, \quad\quad
R^\alpha \ =\ i q^\alpha.   \label{GeneKilling}
\end{eqnarray}
The K\"ahler potential is also given by 
$$
K=\bar q {1\over Q}\log(1+Q) q,
$$
with $Q^\alpha_\beta=-{1\over 2}M^{\beta\delta}_{\alpha\gamma}q^\gamma \bar q_\delta$\cite{Achiman}. 
Then  the Lie-variation of the K\"ahler potential  (\ref{KF}) takes the form 
\begin{eqnarray}
 \varepsilon^A{\cal L}_{R^A} K = i\bar \varepsilon_\alpha q^\alpha -i\varepsilon^\alpha \bar q_\alpha.   \nonumber
\end{eqnarray}
From this we can find the holomorphic functions $F^A$ in (\ref{Gcurr}) to be 
\begin{eqnarray}
 F^A \equiv \{F^\alpha, \bar F_\alpha, F^i, F\}\propto
\{iq^\alpha,0,0,ir\}. \label{F''}
\end{eqnarray}
Here $r$ is a real constant.  There is a classical argument to determine this  $U(1)$ part for the general K\"ahler coset space\cite{Kunitomo}. But at the quantum level we let it be arbitrary as  suggested by the Berkovits method\cite{Berk}.  With  (\ref{G''}), (\ref{GeneKilling}) and (\ref{F''}) 
the currents $J^A \equiv \{J^\alpha, J_\alpha, J^i, J\}$ , proposed by (\ref{Gcurr}), take the form 
\begin{eqnarray}
 J^\alpha &=& {i\over 2}M^{\alpha\beta}_{\gamma\delta}q^\gamma \label{modcurr1}
q^\delta p_\beta  + Aq^\alpha \beta\gamma + C\partial q^\alpha,  \nonumber\\
  J &=& iqp + B\beta\gamma,    \label{modcurr2}\\
 \bar J_\alpha &=& ip_\alpha\quad\quad\quad J^i = ip\Sigma^iq.  \nonumber
\end{eqnarray}
Here $A$, $B$ and $C$ are constants to be determined from the current algebra (\ref{OPE}). We have checked  that these currents indeed satisfy (\ref{OPE}) 
with 
\begin{eqnarray}
  A=-sB, \quad\quad \quad k=i{C\over t} = sN-{1\over s}A^2,   \quad\quad 
  tr \Sigma^i\Sigma^j=k \delta^{ij},  \nonumber   
\end{eqnarray}
and the Killing metric $g^{AB}$ read from the Casimir of the Lie algebra (\ref{geneLie})
$$
\bar X_\alpha X^\alpha + X^\alpha\bar X_\alpha + tH^iH^i +sY^2.
$$

Now we shall find a set of primaries $\phi^I$ with conformal weight $0$ which transforms as the fundamental representation of $G$ 
\begin{eqnarray}
J^A(z)\phi^I(w) \sim {f^{AI}_{\ \ J}\phi^J(w)\over z-w}.  \label{prim}
\end{eqnarray}
Primaries  in any other representation can be constructed by tensoring them. Our conjecture for $\phi^I$ is that 
\begin{eqnarray}
\phi^I = \left(
\begin{array}{c}
  \gamma \phi^{a_1}  \\
\gamma \phi^{a_2} \\
  \vdots \\
    \gamma \phi^{a_n}
\end{array}
\right),    \label{conj}
\end{eqnarray}
in which $\phi^a$'s are holomprphic functions $q^\alpha$ and irreducible components in the decomposition under the subgroup $H$:
\begin{eqnarray}
J^i(z)\phi^a(w) \sim {f^{ia}_{\ \ \ b}\phi^b(w)\over z-w}.   \nonumber
\end{eqnarray}
It is a homogeneous part of the algebra (\ref{prim}). Such irreducible componets can be easily constructed as $J^i$ is given  linearly  in $p_\alpha$ and $q^\alpha$. But they fail to satisfy  the coset part of the algebra (\ref{prim}).
 Here we again have recourse to the Berkovits method. We fermionize the ghost pair as $\beta=\partial \xi e^{-\varphi}$ and $\gamma = \eta e^{\varphi}$\cite{Marti} with
\begin{eqnarray}
\xi(z)\eta(w)\sim {1\over z-w}, \quad\quad\quad 
\varphi(z)\varphi(w) \sim -\log (z-w).   \nonumber
\end{eqnarray}
In (\ref{modcurr1}) and ({\ref{modcurr2}) we replace the quantity $\beta\gamma$ by
\begin{eqnarray}
 \beta\gamma \Rightarrow a\xi\eta + b\partial \varphi,   \label{replace}
\end{eqnarray}
with the constants obeying 
\begin{eqnarray}
a^2-b^2=-1.      \label{ab}
\end{eqnarray}
Then the replacement does not change the OPE
$$
\beta\gamma(z)\cdot\beta\gamma(w)\sim -{1\over (z-w)^2}.
$$
Consequently the $G$-symmetry currents algebra equivalently holds because $\beta\gamma$ in the currents (\ref{modcurr2}) takes part in the algebra through this OPE. However in the algebra (\ref{prim}) it does through $\beta\gamma(z)\cdot\gamma(w)$. 
The fermionization changes this OPE as
\begin{eqnarray}
\beta\gamma(z)\cdot\gamma(w)\Rightarrow (a\xi\eta+b\partial\varphi)(z)\cdot \eta e^\varphi(w) \sim -(a+b){\gamma(w)\over z-w}.    \label{fermi}
\end{eqnarray}
Note that the condition (\ref{ab}) leaves  a freedom still to fix the factor $a+b$ in this equation. We claim that the  algebra (\ref{prim}) with $J^A=J^\alpha, J$ is recovered by choosing it appropriately. Indeed  Berkovits constructed the $SO(10)$ pure spinor $\lambda^\alpha$ on the K\"ahler coset space $SO(10)/U(5)$ in the form  (\ref{pureprim}) by using this fermionization of $\beta$ and $\gamma$\cite{Berk}.   

In this paper we give other examples to support the conjecture (\ref{conj}). The first example is the coset space $SU(n+1)/\{SU(n)\otimes U(1)\}$\cite{Achiman2}. The generators of $SU(n+1)$ are decomposed as $\{X^\alpha,\bar X_\alpha, H^\alpha_{\ \beta}, Y\}$
\begin{eqnarray}
\{X^I_{\ J}\} = \{X^\alpha_{\ n},\bar X^n_{\ \alpha}, X^\alpha_{\ \beta},X^n_{\ n}\} \equiv \{X^\alpha,\bar X_\alpha, H^\alpha_{\ \beta}, Y\}.   \label{gene}
\end{eqnarray}
The Lie algebra takes the form 
\begin{eqnarray}
[ X^\alpha,\bar X_\beta ] &=& H^\alpha_{\ \beta} -(1+{1\over n})\delta^\alpha_{ \beta} Y, \quad\quad\quad [ X^\alpha, X^\beta ] = 0,  \nonumber \\     
  \ [H^\alpha_{\ \beta}, X^\gamma ] &=& \delta^\gamma_\beta X^\alpha -{1\over n}\delta^\alpha_{\beta} X^\gamma, \quad \quad\quad  [Y, X^\alpha ] = - X^\alpha.  
\label{fundU}
\end{eqnarray}
From which we find $M^{\alpha\beta}_{\gamma\delta}=-\delta^\alpha_\gamma \delta^\beta_\delta - \delta^\alpha_\delta \delta^\beta_\gamma$. The quantity $\phi^I$ in the fundamental representation of $SU(n+1)$ transforms as 
\begin{eqnarray}
[ X^\alpha, \left(
\begin{array}{c}
  \phi \\
  \phi^\beta  
\end{array}
\right)  ] &=& \left(
\begin{array}{c}
  \phi^\alpha \\
  0
\end{array}
\right),  \quad\quad\quad       
\ [ X_\alpha, \left(
\begin{array}{c}
  \phi \\
  \phi^\beta 
\end{array}
\right)  ] = 
\left(
\begin{array}{c}
  0 \\
  \delta^\beta_\alpha\phi
\end{array}
\right),            \nonumber\\
\ [ H^\alpha_{\ \gamma}, \left(
\begin{array}{c}
  \phi \\
  \phi^\beta 
\end{array}
\right)  ] &=&  
\left(
\begin{array}{c}
  0 \\
 \delta^\beta_\gamma\phi^\alpha -{1\over n}\delta^\alpha_\gamma\phi^\beta
\end{array}
\right),         \quad\quad
\ [ Y, \left(
\begin{array}{c}
  \phi \\
  \phi^\beta 
\end{array}
\right)  ] = 
\left(
\begin{array}{c}
 {n\over n+1} \phi \ \\
-{1\over n+1} \phi^\beta
\end{array}
\right).     \label{primalgebra}
\end{eqnarray}
The currents corresponding to (\ref{gene}) are given by 
\begin{eqnarray}
J^\alpha &=& -iq^\alpha qp +Aq^\alpha \beta\gamma +C\partial q^\alpha, 
\nonumber \\
J &=& -iqp + B\beta\gamma, \quad\quad
J^\alpha_{\ \beta}= iq^\alpha p_\beta -{i\over n}\delta^\alpha_{\ \beta}qp,  \nonumber \\
\bar J_\alpha &=& ip^\alpha.   \nonumber
\end{eqnarray}
They satisfy (\ref{modcurr2}) with 
$$
A=\sqrt{n+1}, \quad B={n\over \sqrt{n+1}}, \quad iC =1, \quad k= 1,
$$
and the Killing metric $g^{AB}$ taken  from the Casimir of the Lie algebra (\ref{fundU})
$$
\bar X_\alpha X^\alpha + X^\alpha\bar X_\alpha + H^\alpha_{\ \beta}H^\beta_{\ \alpha}  +(1 +{1\over n})Y^2.
$$
We propose the form of the primary fields in the fundamental representation of $SU(n+1)$ as
\begin{eqnarray}
\phi^I = \left(
\begin{array}{c}
  \gamma \\
  \gamma q^\alpha  
\end{array}
\right).
\end{eqnarray}
We find that they satisfy (\ref{prim}) with the coefficients taken from (\ref{primalgebra}) if $-(a+b)A=i$ in the replacement (\ref{replace}).

The second example is the coset space $E_6/\{SO(10)\otimes U(1)\}$\cite{Achiman}. The generators of $E_6$ are decomposed as 
$
\{ X^\alpha,\bar X_\alpha, H^{mn}, Y\}
$
The Lie algebra takes the form 
\begin{eqnarray}
[ H^{mn},H^{kl} ] &=& -i(\delta^{mk}H^{nl} + \delta^{nl}M^{mk} -\delta^{ml}H^{nk}  -\delta^{nk}H^{ml}),   \nonumber\\  
\ [H^{mn}, X^\alpha ] &=& -{i\over 2}(\gamma^{mn})^\alpha_{\ \beta} X^\beta, \quad\quad
 [ Y,X^\alpha ] = -{\sqrt 3\over 2}X^\alpha,  \label{Ealgebra} \\
\ [ X^\alpha, \bar X_\beta ]\ &=& -{i\over 2}(\gamma^{mn})^\alpha_{\ \beta} H^{mn} -\sqrt 3 \delta^\alpha_\beta Y,  \quad\quad  [ X^\alpha,X^\beta ]\ = 0.  \nonumber
\end{eqnarray}
Here $\gamma^{mn}$ are the $SO(10)$ generators in the spinor representation. In the Majorana-Weyl representation the $SO(10)$ Dirac matrices $\Gamma^m$ are given by 
$$
\Gamma^m = \left(
\begin{array}{cc}
  0 & \ \gamma^m \\
 \bar\gamma^m & \ 0
\end{array}
\right).
$$
The Clifford algebra takes the form 
$$
  (\gamma^m\bar \gamma^n +  \gamma^n\bar \gamma^m)^\alpha_{\ \beta} =2\delta^{mn}\delta^\alpha_\beta, \quad\quad
  (\bar \gamma^m\gamma^n + \bar \gamma^n\gamma^m)^\alpha_{\ \beta} =2\delta^{mn}\delta^\alpha_\beta,
$$
and the $SO(10)$ generators are given by either of 
$$
  (\gamma^m\bar \gamma^n -  \gamma^n\bar \gamma^m)^{\ \alpha}_{\beta} =2(\bar \gamma^{mn})^{\ \alpha}_\beta, \quad\quad
  (\bar \gamma^m\gamma^n - \bar \gamma^n\gamma^m)^\alpha_{\ \beta} =2(\gamma^{mn})^\alpha_{\ \beta}.
$$
From (\ref{eq 37}) we find that 
$$
M^{\alpha\beta}_{\gamma\delta} = {1\over 4}(\gamma^{mn})^\alpha_{\ \gamma}(\gamma^{mn})^\beta_{\ \delta} -{3\over 2}\delta^\alpha_{\ \gamma}\delta^\beta_{\ \delta}\ .
$$
The $SO(10)$ currecnts are given by 
\begin{eqnarray}
 J^\alpha &=&{i\over 8}(\gamma^{mn}q)^\alpha p\gamma^{mn}q -{3i\over 4}q^\alpha qp  
 + Aq^\alpha\beta\gamma + C\partial q^\beta,  
\nonumber \\
 J &=& -{\sqrt 3 i\over 2}qp + B\beta\gamma, 
 \quad\quad J_\alpha = ip_\alpha, \nonumber \\
 J^{mn}&=& {1\over 2}p\gamma^{mn}q.    \nonumber 
\end{eqnarray}
We can check that they satisfy the algebra (\ref{OPE}) with
\begin{eqnarray}
 A=2\sqrt 6, \quad B=2\sqrt 2,\quad C=-8i, \quad k =4, \nonumber
\end{eqnarray}
and the Killing metric $g^{AB}$  read from the Casimir of the algebra (\ref{Ealgebra}). 
$$
{1\over 2}(\bar X_\alpha X^\alpha + X^\alpha\bar X_\alpha) + {1\over 2}H^{mn} H^{mn} + Y^2.
$$
The fundamental representation of $E_6$ is ${\bf 27}$, which is decomposed under $SO(10)$ as ${\bf 1}+{\bf 16}+ {\bf 10}$. Correspondingly we have the the primaries  in this representation as  $\phi^I=\{\phi,\phi^\alpha,\phi^k\}$. The transformation law under $E_6$ is given\cite{Kugo2} by
\begin{eqnarray}
[ X^\alpha, \left(
\begin{array}{c}
  \phi \\
  \phi^\beta \\
  \phi^k 
\end{array}
\right)  ] &=& \left(
\begin{array}{c}
  \sqrt 2\phi^\alpha \\
  (\bar\gamma^m)^{\alpha\beta}\phi^m \\
  0
\end{array}
\right),        \nonumber\\
\ [ X_\alpha, \left(
\begin{array}{c}
  \phi \\
  \phi^\beta \\
  \phi^k 
\end{array}
\right)  ] &=& 
\left(
\begin{array}{c}
  0 \\
 \ \sqrt 2\delta_\alpha^\beta\phi \quad  \\
  (\gamma^k \phi)_\alpha
\end{array}
\right),            \nonumber\\
\ [ H^{mn}, \left(
\begin{array}{c}
  \phi \\
  \phi^\beta \\
  \phi^k
\end{array}
\right)  ] &=&  
\left(
\begin{array}{c}
  0 \\
  -{i\over 2}(\gamma^{mn}\phi)^\beta \\
  -i(\delta^{mk}\phi^n-\delta^{nk}\phi^m)
\end{array}
\right),             \label{fundE}   \\
\ [ Y, \left(
\begin{array}{c}
  \phi \\
  \phi^\beta \\
  \phi^k 
\end{array}
\right)  ] &=& \quad\quad
\left(
\begin{array}{c}
 \ {2\over \sqrt 3}\phi \ \\
 \ {1\over 2\sqrt 3}\phi^\beta\ \\
\  -{1\over \sqrt 3}\phi^k  \
\end{array}
\right).     \nonumber
\end{eqnarray}
The Jacobi identity for the transformation law can be checked by means of the fromula
$$
M^{\alpha\beta}_{\gamma\delta}=(\gamma^m)_{\gamma\delta}(\bar\gamma^m)^{\alpha\beta}-2\delta^{[\alpha}_\gamma \delta^{\beta]}_\delta. 
$$
We find that the set of primaries 
\begin{eqnarray}
 \phi^I= \left(
\begin{array}{c}
 {1\over \sqrt 2} \gamma   \\
 \gamma q^\alpha \\
  {1\over 2}\gamma q\gamma^m q  
\end{array}
\right) 
\end{eqnarray}
satisfies the algebra (\ref{prim}) with the coefficients $f^{AI}_{\ \ J}$ taken from (\ref{fundE}) if $-(a+b)A=2i$ in the replacement (\ref{replace}). This expression itself  was found in \cite{Nitta} without quantization for discussing other physics.

To conclude our arguments some comments are in order. 
The form of the $G$-symmetry currents in (\ref{Gcurr}) was inferred from the Killing potentials
\begin{eqnarray}
-iM^A=K_{,\alpha}R^{A\alpha} -F^A, \nonumber
\end{eqnarray}
which exist for the general K\"ahler coset space and satisfy
 the Lie algebra of $G$
\begin{eqnarray}
 {\cal L}_{R^A}M^B = f^{ABC}M^C.   \label{LieKilling}
\end{eqnarray}
Here $K$ is the K\"ahler potential and   $F^A$ are the holomorphic functions   $F^A$  found from  (\ref{KF})\cite{bagger}. 
The K\"ahler $2$-form (\ref{metric'}) can be put in the form 
\begin{eqnarray}
\omega =d p_\alpha\wedge d q^\alpha,    \nonumber
\end{eqnarray}
with $p_\alpha=iK_{,\alpha}$. The Killing potentials $M^A$ are rewritten in terms of the so-called Darboux coordinates $(p_\alpha,q^\alpha)$\cite{MA1} as
\begin{eqnarray}
M^\alpha &=& {i\over 2}M^{\alpha\beta}_{\gamma\delta}q^\gamma q^\delta p_\beta +q^\alpha,   \quad\quad\quad   \bar M_\alpha =ip_\alpha,  \nonumber \\
 M^i&=& ip\Sigma^i q, \quad\quad\quad\quad M=ipq -1.   \nonumber
\end{eqnarray}
Here the $U(1)$ part in $F^A$, which was  left arbitrary as $r$ in (\ref{F''}),  was fixed  so that  
(\ref{LieKilling}) is fulfilled. This way of fixing the $U(1)$ part\cite{bagger} is alternative to the one discussed in \cite{Kunitomo} for the general K\"ahler coset space. It is now clear that  there is a close relationship  between these Killing potentials and the $G$-symmetry currents (\ref{modcurr2}). 
In \cite{MA1} it was discussed that the introduction of the Darboux coordinates simplifies quantum deformation of the K\"ahler manifold. Namely quantum deformation of the symplectic manifold defined by (\ref{metric}) is done through the non-commutative $\star$ product 
$$
f(x)\star g(x) = \sum_n {1\over n!}\Big(-{i\hbar\over 2}\Big)^n \omega^{i_1j_1}
\omega^{i_2j_2}\cdots \omega^{i_nj_n} \partial_{i_1}\partial_{i_2}\cdots f(x)
 \partial_{j_1}\partial_{j_2}\cdots \partial_{j_n}g(x),
$$
according to Fedosov\cite{Fed}, in which  $\omega^{ij}=g^{ik}g^{jl}\omega_{kl}$. When the symplectic manifold is k\"ahlerian,  the $\star$ product becomes the well-known  Moyal product by using the Darboux coordinates. Then it is easy to study  the non-commutative algebrae for the Killing potentials
\begin{eqnarray}
[ M^A(q,\bar q),M^B(q,\bar q) ]_\star &=& -i(c_1\hbar + c_3\hbar^3 + c_5\hbar^5+\cdots )  f^{ABC}M^C,  \nonumber\\
 M^A(q,\bar q)\star M^A(q,\bar q) &=& c_0 + c_2\hbar^2 + c_4\hbar^4+ \cdots.
\nonumber
\end{eqnarray}
In \cite{MA1,MA2}  the numerical constants $c_j,(j\ge0)$  were found to be 
\begin{eqnarray}
c_0&=&R \ \  ({\rm Riemann\ scalar }),  \quad c_1=1,  \nonumber \\
 c_2&=& -{1\over 2}(tr \Sigma^i\Sigma^i + N),  \quad\quad\quad
c_i=0 \quad {\rm for}\ \   i\ge 3.     \nonumber
\end{eqnarray}
 The Berkovits method discussed in this letter indicates that the fuzzy K\"ahler coset space 
may be studied  by incoorporating the bosonic ghosts and  generalizing the the Moyal product  
\begin{eqnarray}
&\quad &f(p,q,\beta,\gamma)\star g(p,q,\beta,\gamma)    \nonumber \\
 &=& \sum_n {1\over n!}f(p,q,\beta,\gamma)\Big[ -{i\hbar\over 2}
\Big({\overrightarrow \partial\over \partial p_\alpha}{ \overleftarrow\partial\over \partial q^\alpha}+{\overrightarrow \partial\over \partial \beta}{\overleftarrow \partial\over \partial \gamma} -{\overleftarrow \partial\over \partial p_\alpha}{\overrightarrow \partial\over \partial q^\alpha}  - {\overleftarrow \partial\over \partial \beta}{\overrightarrow \partial\over \partial \gamma} \Big)\Big]^n f(p,q,\beta,\gamma).  \nonumber
\end{eqnarray}
The study in this direction is expected to shed a new light on the non-commutative geometry of the K\"ahler coset space. 

In this letter the $G$-symmetry currents have been explicitly given for the irreducible  K\"ahler coset space $G/H$ as (\ref{modcurr2}). The $G$-symmetry primaries could not be given in such a general form because the decomposition (\ref{conj}) under the subgroup $H$ varies from case to case. But we are sure of being able to construct  them explicitly for other types of the irreducible K\"ahler coset space. The construction may be extended to the reducible K\"ahler coset space $G/\{S\otimes U(1)^k\}$\cite{Kunitomo,MA3,A}, of which extreme case $G/U(1)^r$ is the flag manifold used for the Wakimoto realization of the Lie algebra of $G$. The study is undergoing. 

\vspace{1cm}
\noindent
\noindent
{\Large\bf Acknowledgements}

The work was supported in part  by the Grant-in-Aid for Scientific Research No.
13135212.

\end{document}